\documentclass{article}
\usepackage{macros/arxiv}

\usepackage{amsfonts}
\usepackage{amssymb}
\usepackage{amsbsy} 
\usepackage{amsmath}
\usepackage{cuted}
\usepackage{flushend}
\usepackage{caption}
\usepackage{subcaption}
\usepackage[dvips]{graphicx}
\usepackage[T1]{fontenc}
\usepackage[latin1]{inputenc}
\usepackage[table,xcdraw]{xcolor}

\usepackage{hyperref}
\hypersetup{
    unicode=false,          
    pdftoolbar=true,        
    pdfmenubar=true,        
    pdffitwindow=false,     
    pdfstartview={FitH},    
    pdftitle={IEEE submission},    
    pdfauthor={Prof.Fassman},     
    pdfsubject={RBIG-Information},   
    pdfcreator={Prof.Fassman},   
    pdfproducer={Prof.Fassman}, 
    pdfkeywords={keyword1} {key2} {key3}, 
    pdfnewwindow=true,      
    colorlinks=true,       
    linkcolor=blue,          
    citecolor=blue,        
    filecolor=blue,      
    urlcolor=blue           
}

\usepackage{wrapfig,booktabs}

\title{\textit{RotNet}: Fast and Scalable Estimation of Stellar Rotation Periods Using Convolutional Neural Networks}

\author{%
   J. Emmanuel Johnson \\
   Universit\"at de Val\`encia \\
   \texttt{juan.johnson@uv.es} \\
   \And
   Sairam Sundaresan \\
   Intel Labs\\
   \texttt{sairam.sundaresan@intel.com} \\
   \And
   Tansu Daylan \\
   Massachusetts Institute of Technology \\
   \texttt{tdaylan@mit.edu} \\
   \And
   Lisseth Gavilan \\
   NASA Ames Research Center \\
   \texttt{lisseth.gavilanmarin@nasa.gov} \\
   \And
   Daniel K. Giles \\
   Illinois Institute of Technology \\
   \texttt{dgiles1@hawk.iit.edu} \\
   \And
   Stela Ishitani Silva \\
   The Catholic University of America \\
   \texttt{ishitanisilva@cua.edu} \\
   \And
   Anna Jungbluth \\
   University of Oxford \\
   \texttt{anna.jungbluth@physics.ox.ac.uk}
   \And
   Brett Morris \\
   University of Bern \\  
   \texttt{brett.morris@space.unibe.ch} \\
   \And
   Andr\'es Mu\~noz-Jaramillo \\
   Southwest Research Institute\\
   \texttt{amunozj@boulder.swri.edu} \\
}

\begin{document}


\maketitle

\begin{abstract}


Magnetic activity in stars manifests as dark spots on their surfaces that modulate the brightness observed by telescopes. These light curves contain important information on stellar rotation. However, the accurate estimation of rotation periods is computationally expensive due to scarce ground truth information, noisy data, and large parameter spaces that lead to degenerate solutions.
%
%
%
We harness the power of deep learning and successfully apply Convolutional Neural Networks to regress stellar rotation periods from \textit{Kepler} light curves. Geometry-preserving time-series to image transformations of the light curves serve as inputs to a ResNet-18 based architecture which is trained through transfer learning. The McQuillan catalog of published rotation periods is used as ansatz to ground truth.
We benchmark the performance of our method against a random forest regressor, a 1D CNN, and the Auto-Correlation Function (ACF) - the current standard to estimate rotation periods. Despite limiting our input to fewer data points ($\sim$1k), our model yields more accurate results and runs 350 times faster than ACF runs on the same number of data points and 10,000 times faster than ACF runs on $\sim$65k data points. With only minimal feature engineering our approach has impressive accuracy, motivating the application of deep learning to regress stellar parameters on an even larger scale. 

\end{abstract}
\newpage
\section{Introduction}
    \label{sec:intro}
    \begin{wrapfigure}{R}{0.30\textwidth} 
\vspace{-20pt}
\begin{center}
\includegraphics[width=4cm]{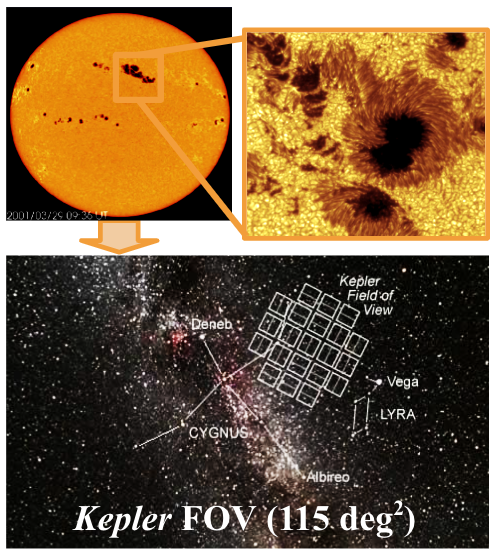}
\caption{Spots occur on billions of stars in Kepler's FOV.} 
\label{fig:keplerfov}
\end{center}
\vspace{-10pt}
\end{wrapfigure}
Magnetic activity forms dark spots on the surface of stars. Through telescopes, these spots are only visible as slight dips in the brightness measured over time (light curves).
Parameterizing spot properties (like their number, size, and location) is crucial to understanding stellar magnetic activity but also very challenging. Spots move on the stellar surface, and many parameter combinations can lead to the same 1D signal - a problem known as degeneracy. The easiest parameter to determine, and one critical to constrain others, is the stellar rotation period, P$_{Rot}$.



Different methods are used to study rotation in photometric light curves. One class of methods involves feature-engineering whereby clever transformations of light curves are used to better represent the information they contain. These include detecting peaks in Lomb-Scargle (LS) periodograms (e.g. \cite{Reinhold2013, Vida2018}), Auto-Correlation Functions (ACFs) (e.g. \cite{McQuillan2013}), and time-frequency analysis such as wavelet transforms (e.g. \cite{Mathur2010, Garcia2014}). The ACF method in particular has proven extremely effective for measuring rotation periods. The most comprehensive catalog of rotation periods for the Kepler dataset utilized the ACF method, finding it more robust to noise, systematics, and evolution of stellar activity than periodograms \cite{McQuillan2013, McQuillan2014}.
Another class of algorithms rely on data-driven approaches popular in ML. Neural Networks and Random Forests (RF) have been leveraged to predict if a light curve will produce a robust rotation period measurement from its Lomb-Scargle periodogram \cite{Agueros2018}. 
More recently, \cite{Breton2019} used RF to classify rotating and non-rotating stars and trained a second RF to find the best high-pass filter to retrieve the most probable rotation period.

%

Here we present a concept study demonstrating that pre-trained Convolutional Neural Networks can estimate stellar rotation periods from thousands of \textit{Kepler} light curves, covering a large field of view (FOV) of the Galaxy (Fig. \ref{fig:keplerfov}). This new pipeline offers similar or better mean absolute errors (MAE) to the ACF method, but is orders of magnitudes faster while only requiring a fraction of the data.

\section{Methods}
    \label{sec:methods}
    \subsection{Data}
In this work, we use data from the \textit{Kepler} Mission. The \textit{Kepler} Space Telescope observed $>$200k targets for four years. Here, we use long cadence light curves, i.e. time series data which recorded the brightness of a target every 30 minutes. The \textit{Kepler} spacecraft was in an earth-trailing orbit around the sun, and every three months the spacecraft rotated to maximize the solar panel efficiency. As such, the data is separated into quarters (Q) of roughly 90 days.
The continuous four-year, high precision photometry enabled precise determination of stellar rotation periods for tens of thousands of stars. McQuillan et al. produced a catalog of rotation periods using the ACF method applied to continuous light curves from Q3-Q14, and continuous three quarter segments (Q3-Q5, Q6-Q8, etc.), and reported rotation periods where there was agreement in period determination. This redundancy gives high confidence of accuracy to their published catalog.

\subsection{Baselines}
We establish ACF and Machine Learning (ML) baselines as precursors to the final work. The ACF method represents the astrophysics standard, and the ML models start from common approaches and build in complexity.

\textbf{ACF.} To establish comparison on computational cost, time, and accuracy, we use the ACF method to determine the rotation periods for over 100 thousand Kepler targets. We use all 17 quarters of Kepler data, encompassing four years of observations, to produce the best possible results. Additionally, we use the ACF approach on single quarters of data (three months, and 1080 data points) for a direct performance comparison to our ML approaches (Fig. \ref{fig:results}).

\textbf{Machine Learning Baselines.}
We first investigated the performance of simple Random Forests and 1D Convolutional Neural Networks (CNNs) on the data. 
Our model architecture choices ranged from a simple two layer CNN to more complicated designs with several convolution, pooling and activation layers appended to a regression module. In order to ensure robust baselines using these architectures, we performed extensive hyperparameter sweeps which is detailed in the experiments section.

Through these baseline experiments, we observed that the regressed rotation periods from each baseline had high mean absolute error (MAE). Further, the 1D CNN models struggled to converge. This highlights the challenges of significant degeneracy in the data. Given these results, we transitioned to 2D CNNs by mapping our 1D time series into images.


\subsection{RotNet}
In recent times, CNNs \cite{NIPS2012} have been the default building blocks for modern computer vision applications. Further, models trained on large datasets can be extended to solve new problems via transfer learning \cite{DLTrans}. Given the inherent complexity in our data, and the limited amount of reliable labels (estimates from McQuillan et. al), we map the light curves into images (detailed in Section \ref{sec:img_trans}) and then harness the power of transfer learning to regress stellar rotation periods. 

\subsubsection{Overall Pipeline}
\begin{figure}[h!]
\small
\begin{center}
\includegraphics[width=14.0cm]{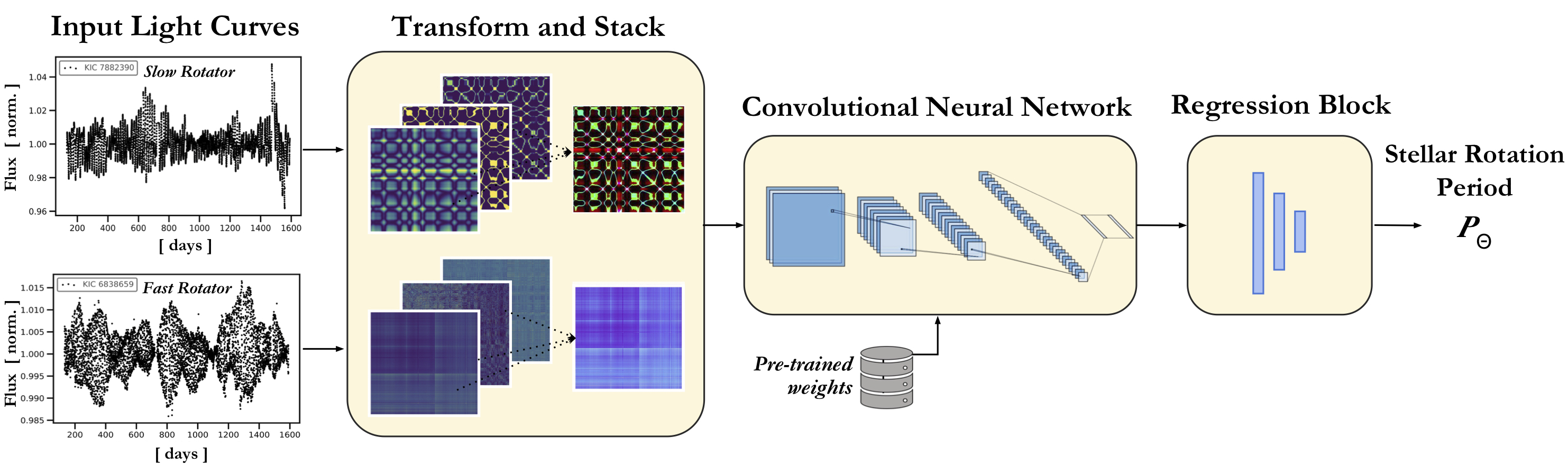}
\caption{\textit{RotNet} pipeline: Given a light curve (spanning slow to fast rotators) we apply three image  transforms and stack them as channels. The stacked images are then fed to a CNN feature extractor which is initialized with pre-trained weights. The features thus obtained enable the regression block to produce estimates of the desired stellar rotation period  P$_{\theta}$.}
\label{fig:pipeline}
\end{center}
\end{figure}

The overall pipeline, which we call \textit{RotNet}, is shown in Fig. \ref{fig:pipeline}. Given a light curve as input, we first transform it into a three channel image. Each of these transforms yields an $N \times N$ image channel given a light curve of $N$ time steps. We employ transfer learning by initializing our CNN with pre-trained weights from the ImageNet dataset \cite{russakovsky2014imagenet}. We remove the classification head from the CNN and instead append a regression block to produce rotation period estimates. Here, we choose ResNet-18 \cite{resnet18, resnet182} to act as the backbone feature extractor for the imaged time series signal and these features are fed into the regression block which consists of fully connected layers with dropout and ReLU activation functions. 
The key enabler of our approach is the Time Series-to-Image transformations for which we explain each of the steps in the next section. 

\subsubsection{Image Transformations} \label{sec:img_trans}
Prior work  \cite{Wang2015, Eckmann_1987} has successfully demonstrated the use of CNNs on transformed time series data. We use these studies as inspiration. A vast majority of CNNs use images with three channels as input. Hence we stack three different transformations to form an image that can be processed by a CNN. Concretely, we use the Gramian Angular Field, Markov Transition Field, and Recurrence Plot transforms. Each of these transformations preserve the temporal correlation in the data, while enabling the CNN to extract rich and meaningful features for regressing properties of the light curve. 

The \textbf{Gramian Angular Field (GAF)} is gram matrix representation of the inner product between each point within the time series \cite{Wang2015}. A non-linear function is called to convert the coordinates into the polar coordinate system. This transformation preserves the temporal geometry and is fairly robust to noise.   
The \textbf{Markov Transition Field (MTF)} is another gram matrix representation with the Markov transition probabilities preserving the information in the time domain \cite{Wang2015} .
Finally, the \textbf{Recurrence Plot (RP)} is constructed by representing the distance between trajectories from the original time series \cite{Eckmann_1987}. Once a simple gram matrix via the euclidean distance between the trajectories is calculated, a threshold is applied to only take the most influential points.

The transformations were computed using PyTs \cite{JMLRpyts}, which is a fast time series library. Although this increases the computational burden by $\mathcal{O}(N^2)$ data points in memory, it allows us to leverage established computer vision architectures to effectively deal with image representations.


\section{Experiments} \label{sec:train_dets}  
\subsection{Data Preparation}  We ran ACF on $100$ thousand Kepler targets and we sorted the light curves by rotation periods and selected $18,472$ of the fastest rotating light curves each with $1,080$ time steps. We then 
split the data into a training set ($70\%$), test set ($20\%$), and validation set ($10\%$). This resulted in $14,439$ light curves in the training set, $2,047$ in the validation set, and $4,186$ in the testing set. More importantly, we ensured that the distribution of rotation periods was the same for all three splits. The light curves (inputs) and rotation periods (outputs) were zero centered with unit standard deviation. As seen in figure \ref{fig:results}, most of the rotation periods were smaller, so to account for this, we composed a log-transformation, a quantile transformation and min-max scaling to center the data in the interval $[0,1]$. This particularly helped with training the CNNs. 

\subsection{Hyperparameters}
\textbf{Random Forest.} Using grid search and five fold cross validation, we performed a sweep to determine the optimal parameters for the model. The search space included variations in the number of estimators, the maximum number of features to obtain the best split, and the minimum samples required to split a node.  

 \textbf{CNN.} Our 1D and 2D CNN architectures were trained using the Adam optimizer and a multi-step learning rate schedule with drops at $30$, $60$ and $80$ percent of the total epochs and a decay factor of $0.1$. Using standardized experimental setups with \textit{PyTorch-Lightning} \cite{falcon2019pytorch}, we ran hyper-parameter sweeps using the \textit{Weights and Biases} tool \cite{wandb} over a wide range of parameters including batch size, dropout, learning rate as well as the model architecture (Custom 1D CNNs, ResNet-18, ResNet-50, Vgg-16). In the case of the 2D CNNs, we also evaluated training the models from randomly initialized weights. However, we found that initializing the models using pretrained weights significantly accelerated convergence. Both the mean squared error and mean absolute error were considered as loss functions in the sweeps. We obtained our best results with a batch size of 16, a dropout of 0.1, an initial learning rate of 0.001 with the 1D CNN and the same hyperparameters with a pretrained ResNet-18 backbone for the 2D CNN.

\section{Results}
    \label{sec:results}
    \begin{figure}[t!]
\small
\begin{center}
\includegraphics[width=\textwidth]{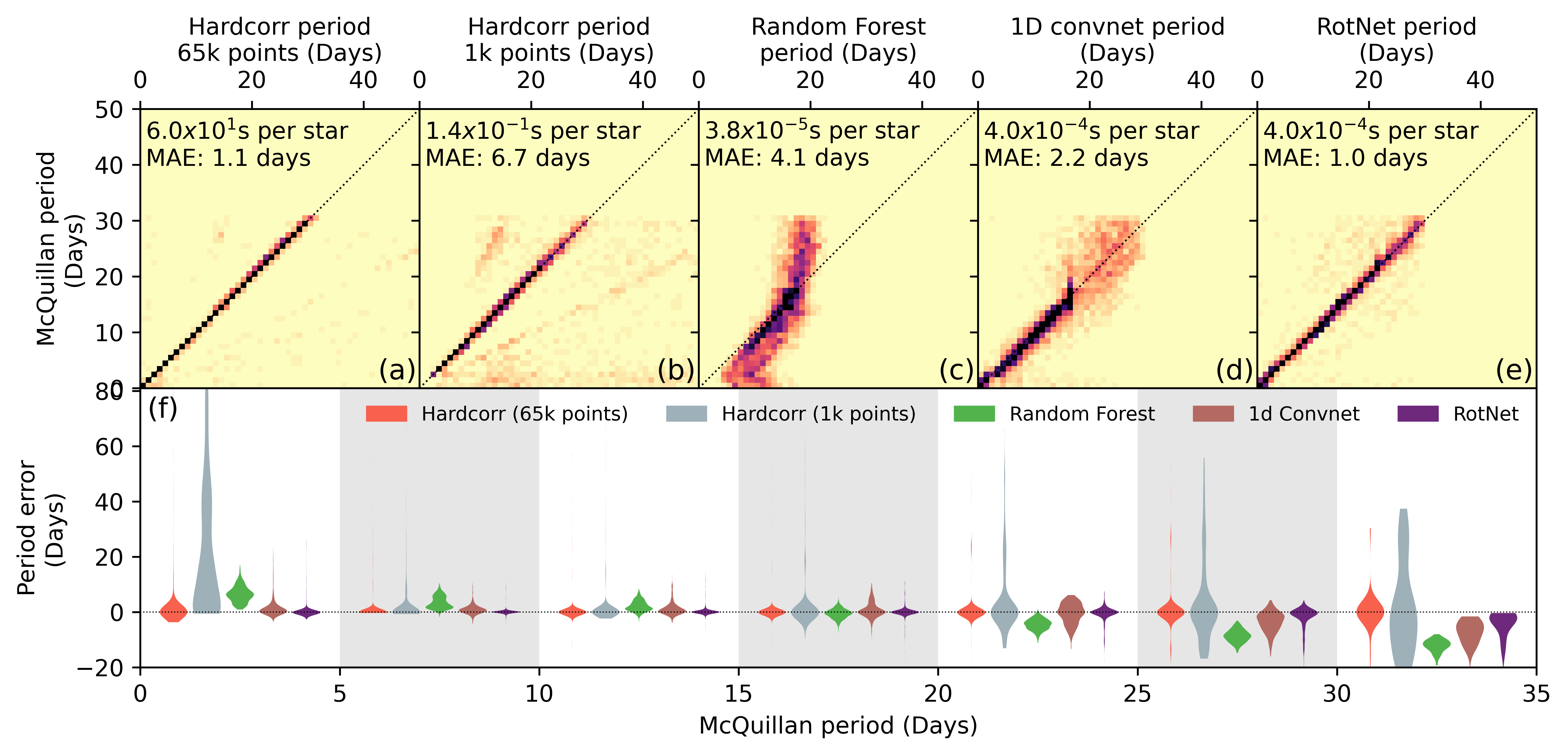}
\caption{Performance of period detection algorithms in comparison to the McQuillan catalog \cite{McQuillan2014}. \textit{Top row}: 2D histogram of calculated periods for the 3,700 stars in our test set vs. the period calculated by McQuillan.  The histogram uses uniform vertical and horizontal spatial grids with one day intervals.  Each panel states how long does it take to process a stellar lightcurve in a single CPU processor.  It also displays the Mean Absolute Error (MAE) for all stars in the set.  \textit{Bottom-row}: Violin plots displaying the distribution of residual errors for different 5 degree period bins (each bin is represented using a white or gray shaded area).  Violin plots for different techniques are offset horizontally in each bin, but are built using the same subset of McQuillan stars.} 
\label{fig:results}
\end{center}
\end{figure}

Figure \ref{fig:results}-a shows that our ACF implementation yields results that are consistent with the McQuillan catalog.
These results are obtained using $17$ Kepler quarters at a $30$ minute cadence ($\approx65k$ data points per light curve). This process takes $60$ seconds per star on a single CPU core.  The computational cost of ACF can be reduced by reducing the number of data points used, but not without incurring a significant drop in accuracy.  As shown in Figure \ref{fig:results}-b, reducing the light-curve to $\approx1$k points is $100$x faster, but results in a significant performance hit. Using the baseline Random Forest model, we were able to achieve a significant speedup as can be seen in \ref{fig:results}-c. However, the fit is qualitatively worse even though it is quantitatively better than the ACF trained on $1$k stars in terms of Mean Absolute Error (MAE). On the other hand, the trained CNNs performed better in terms of accuracy \textit{and} speed than our baseline Random Forest and ACF.  Figure \ref{fig:results}-d shows the results when using our 1D CNN and Figure \ref{fig:results}-e shows the results of RotNet after transforming the light-curves to images.  There are two clear advantages of using CNNs:  1.\ Inference is now $\approx10,000$x faster than the full ACF approach for both 1D and 2D CNNs.  2.\ Our proposed method, RotNet, has a similar performance to the full ACF approach even though it uses only $\approx1$k data points in each light curve, as opposed to $\approx65$k in the full ACF solution.  Figure \ref{fig:results}-f (second row) shows the distribution of errors for different period bins.  Our RotNet solution is closer to the McQuillan results than any of our other methods and shows the tightest error distribution.  It clearly outperforms the reduced $1\mathcal{k}$ ACF method which struggles to recover the period of fast rotators (0-5 days) and tends to underestimate slow rotators (30-35 days). Note that we have limited our analysis to stars with P$_{Rot}$ $\leq$ 30 days due to their high abundance and signal to noise characteristics.


\section{Conclusions / Future Work}
    \label{sec:conclusion}
    Exoplanet hunting missions and surveys (i.e.\ \textit{Kepler} and \textit{TESS}) have already generated Terabytes of stellar light curves that are a treasure trove of data for understanding exoplanet hosts, stellar rotation and magnetism.  However, traditional algorithms used to estimate stellar properties, like ACF, are expensive and require long observational baselines.  Here we demonstrate that our pipeline (RotNet) based on a supervised, pre-trained Convolutional Neural Networks is able to estimate stellar rotation periods with a similar level of accuracy as the full ACF approach. It does so with 65 times less data points and is 10,000 times faster. Training and deploying CNNs requires the existence of a ground truth, which we assume to be the period estimates obtained by McQuillan et al.  This means that RotNet is only as good as the method used to determine its training target stellar properties.   Nevertheless, our results demonstrate that neural networks can be used to massively scale methods that are otherwise expensive and can only be applied to a handful of stars.  On its own, RotNet is not meant to be a substitution for more detailed inference methods of stellar properties.  Instead, the combination of direct methods of inference with supervised machine learning paints a bright future to fully take advantage of current and future missions that measure the time evolution of stellar light curves.  In the future, we plan to extend this framework to other stellar properties like inclination, before tackling stellar magnetic parameters.

\section*{Broader Impact}
    \label{sec:impact}
There are no ethical or societal consequences of our work except that it allows our Sun, its magnetic properties, and space weather to be placed in cosmological context. Our work, however, can significantly benefit the astronomical community by characterizing the rotation of millions of stars in near-future high-cadence sky surveys such as that will be carried out by the Rubin Observatory, which would not be computationally feasible with forward-modeling. This can then lead to the characterization of the magnetic properties of these stars with a wide range of radii, masses, metallicities, effective temperatures, and ages.

Currently, there are a few methods by which magnetism on other stars can be probed. Polarized line emission and absorption (i.e., Doppler-Zeeman imaging) has been used extensively to map the magnetic field lines on bright stars. Second, high-resolution imaging via optical interferometers, such as the Center for High Angular Resolution Astronomy (CHARA) \cite{Parks2011}, are used to spatially resolve stars and characterize their spots. The disadvantage of these probes is that the sample size is limited to the nearest and brightest stars, which precludes generalizations of conclusions to arbitrary stars. Our methodology unifies and extends the currently limited census of star spots by providing a robust scheme to interpolate known relations in stellar magnetism as well as opening up the potential to discover new ones. This gives an observational probe to determine spot seasons, latitudinal distribution of star spots, and spot coverage. It also allows us to study how stellar magnetic activity is related to stellar properties such as radius, mass, metallicity, effective temperature, and age. In turn, this characterization leads to a better understanding of stellar dynamo theory.

Stellar rotation is inferred thanks to the existence of star spots. Star spots can rotate with different angular velocities at different latitudes. Spots that rotate with similar, but distinct angular velocities can therefore generate beating patterns and cause star spot-induced light curve features to evolve at long time scales. However, star spots \emph{also} physically evolve over such time scales by changing their contrast, shape, or size. Therefore, a well-known degeneracy in this inference problem is that of distinguishing star spots subject to differential rotation and evolution. Beyond the work presented here, our research aims to address this question by incorporating physical generative models as well as exploiting the long observation baseline of the \textit{Kepler} telescope.

Ideally one would like to construct a machine learning pipeline that can predict the star spot maps of any star with an arbitrary set of parameters. However, this is currently not possible, because our training data are constructed using photometric measurements of a particular survey (e.g., \textit{Kepler}, \textit{TESS}), which has an inherent target selection bias. The failure to characterize potential target selection biases or other biases due to systematic errors in the data would lead to spurious inferences.


\section*{Acknowledgements}


This work was initiated at the NASA Frontier Development Lab (FDL) 2020. NASA FDL is a public-private partnership between NASA, the SETI Institute and private sector partners including Google Cloud, Intel, IBM, and NVIDIA, amongst others. These partners provide the data, expertise, training, and compute resources necessary for rapid experimentation and iteration in data-intensive areas. We would like to thank Weights \& Biases for their support and for providing us with an academic license to use their experiment tracking tools for our work.

The authors sincerely acknowledge the invaluable discussions with mentors of the FDL 2020 program: Yarin Gal (Oxford), Gibor Basri (UC Berkeley), Antonino Lanza (INAF), and Valentina Salvatelli (IQVIA).  
JEJ acknowledges support from European Research Council.
TD acknowledges support from MIT's Kavli Institute as a Kavli postdoctoral fellow.
LG is supported by a NASA postdoctoral program (NPP) fellowship.
DG thanks the LSSTC Data Science Fellowship Program, which is funded by LSSTC, NSF Cybertraining Grant \#1829740, the Brinson Foundation, and the Moore Foundation; his participation in the program has benefited this work.

\bibliographystyle{plain}
\bibliography{refs}

\begin{thebibliography}{10}

\bibitem{Agueros2018}
Marcel {Ag{\"u}eros} and Alexander {Teachey}.
\newblock {Using Machine Learning To Predict Which Light Curves Will Yield
  Stellar Rotation Periods}.
\newblock In {\em American Astronomical Society Meeting Abstracts \#231},
  volume 231 of {\em American Astronomical Society Meeting Abstracts}, page
  349.21, January 2018.

\bibitem{wandb}
Lukas Biewald.
\newblock Experiment tracking with weights and biases, 2020.
\newblock Software available from wandb.com.

\bibitem{Breton2019}
S.~N. {Breton}, L.~{Bugnet}, A.~R.~G. {Santos}, A.~{Le Saux}, S.~{Mathur},
  P.~L. {Pall{\'e}}, and R.~A. {Garc{\'\i}a}.
\newblock {Determining surface rotation periods of solar-like stars observed by
  the Kepler mission using machine learning techniques}.
\newblock In P.~{Di Matteo}, O.~{Creevey}, A.~{Crida}, G.~{Kordopatis},
  J.~{Malzac}, J.~B. {Marquette}, M.~{N'Diaye}, and O.~{Venot}, editors, {\em
  SF2A-2019: Proceedings of the Annual meeting of the French Society of
  Astronomy and Astrophysics}, page~Di, December 2019.

\bibitem{Eckmann_1987}
J.-P Eckmann, S.~Oliffson Kamphorst, and D~Ruelle.
\newblock Recurrence plots of dynamical systems.
\newblock {\em Europhysics Letters ({EPL})}, 4(9):973--977, nov 1987.

\bibitem{falcon2019pytorch}
WA~Falcon.
\newblock Pytorch lightning.
\newblock {\em GitHub. Note:
  https://github.com/PyTorchLightning/pytorch-lightning}, 3, 2019.

\bibitem{JMLRpyts}
Johann Faouzi and Hicham Janati.
\newblock pyts: A python package for time series classification.
\newblock {\em Journal of Machine Learning Research}, 21(46):1--6, 2020.

\bibitem{Garcia2014}
R.~A. {Garc{\'\i}a}, T.~{Ceillier}, D.~{Salabert}, S.~{Mathur}, J.~L. {van
  Saders}, M.~{Pinsonneault}, J.~{Ballot}, P.~G. {Beck}, S.~{Bloemen}, T.~L.
  {Campante}, G.~R. {Davies}, Jr. {do Nascimento}, J.~D., S.~{Mathis}, T.~S.
  {Metcalfe}, M.~B. {Nielsen}, J.~C. {Su{\'a}rez}, W.~J. {Chaplin},
  A.~{Jim{\'e}nez}, and C.~{Karoff}.
\newblock {Rotation and magnetism of Kepler pulsating solar-like stars. Towards
  asteroseismically calibrated age-rotation relations}.
\newblock {\em Astronomy and Astrophysics}, 572:A34, December 2014.

\bibitem{resnet182}
K.~{He}, X.~{Zhang}, S.~{Ren}, and J.~{Sun}.
\newblock Deep residual learning for image recognition.
\newblock In {\em 2016 IEEE Conference on Computer Vision and Pattern
  Recognition (CVPR)}, pages 770--778, 2016.

\bibitem{NIPS2012}
Alex Krizhevsky, Ilya Sutskever, and Geoffrey~E Hinton.
\newblock Imagenet classification with deep convolutional neural networks.
\newblock In F.~Pereira, C.~J.~C. Burges, L.~Bottou, and K.~Q. Weinberger,
  editors, {\em Advances in Neural Information Processing Systems 25}, pages
  1097--1105. Curran Associates, Inc., 2012.

\bibitem{Mathur2010}
S.~{Mathur}, R.~A. {Garc{\'\i}a}, C.~{R{\'e}gulo}, O.~L. {Creevey},
  J.~{Ballot}, D.~{Salabert}, T.~{Arentoft}, P.~O. {Quirion}, W.~J. {Chaplin},
  and H.~{Kjeldsen}.
\newblock {Determining global parameters of the oscillations of solar-like
  stars}.
\newblock {\em Astronomy and Astrophysics}, 511:A46, February 2010.

\bibitem{McQuillan2013}
A.~{McQuillan}, S.~{Aigrain}, and T.~{Mazeh}.
\newblock {Measuring the rotation period distribution of field M dwarfs with
  Kepler}.
\newblock {\em mnras}, 432(2):1203--1216, June 2013.

\bibitem{McQuillan2014}
A.~{McQuillan}, T.~{Mazeh}, and S.~{Aigrain}.
\newblock {Rotation Periods of 34,030 Kepler Main-sequence Stars: The Full
  Autocorrelation Sample}.
\newblock {\em The Astrophysical Journal Supplement Series}, 211(2):24, April
  2014.

\bibitem{DLTrans}
S.~J. {Pan} and Q.~{Yang}.
\newblock A survey on transfer learning.
\newblock {\em IEEE Transactions on Knowledge and Data Engineering},
  22(10):1345--1359, 2010.

\bibitem{Parks2011}
J.~R. Parks, R.~J. White, G.~H. Schaefer, J.~D. Monnier, and G.~W. Henry.
\newblock Starspot imaging with the chara array.
\newblock In Christopher Johns-Krull, Matthew~K. Browning, and Andrew~A. West,
  editors, {\em 16th Cambridge Workshop on Cool Stars, Stellar Systems, and the
  Sun}, volume 448 of {\em Astronomical Society of the Pacific Conference
  Series}, page 1217, December 2011.

\bibitem{Reinhold2013}
Timo {Reinhold}, Ansgar {Reiners}, and Gibor {Basri}.
\newblock {Rotation and differential rotation of active Kepler stars}.
\newblock {\em Astronomy and Astrophysics}, 560:A4, December 2013.

\bibitem{russakovsky2014imagenet}
Olga Russakovsky, Jia Deng, Hao Su, Jonathan Krause, Sanjeev Satheesh, Sean Ma,
  Zhiheng Huang, Andrej Karpathy, Aditya Khosla, Michael Bernstein,
  Alexander~C. Berg, and Li~Fei-Fei.
\newblock Imagenet large scale visual recognition challenge, 2014.

\bibitem{Vida2018}
Kriszti{\'a}n {Vida} and Rachael~M. {Roettenbacher}.
\newblock {Finding flares in Kepler data using machine-learning tools}.
\newblock {\em Astronomy and Astrophysics}, 616:A163, September 2018.

\bibitem{Wang2015}
Zhiguang Wang and T.~Oates.
\newblock {Encoding time series as images for visual inspection and
  classification using tiled convolutional neural networks}.
\newblock In {\em AAAI Workshop - Technical Report}, 2015.

\bibitem{resnet18}
S.~{Xie}, R.~{Girshick}, P.~{Dollár}, Z.~{Tu}, and K.~{He}.
\newblock Aggregated residual transformations for deep neural networks.
\newblock In {\em 2017 IEEE Conference on Computer Vision and Pattern
  Recognition (CVPR)}, pages 5987--5995, 2017.

\end{thebibliography}
\end{document}